\documentclass[12pt,thmsa]{article}
 \usepackage{amsmath}

 \usepackage{amssymb}

\input{tcilatex}

\begin{document}

\baselineskip 17pt

\noindent {\large {\bf RELATIVISTIC LENGTH AGONY CONTINUED}}

\bigskip

\centerline {\bf Dragan V Red\v zi\' c}

\noindent {\small Faculty of Physics, University of Belgrade, PO Box
368, 11001 Beograd, Serbia, Yugoslavia}\\ E-mail address:
redzic@ff.bg.ac.rs

\bigskip

\noindent {\bf Abstract}

\noindent An attempt is made to remedy confusing treatments of some
basic relativistic concepts and results in recent papers by Franklin
(2010 {\it Eur. J. Phys.} {\bf 31} 291-8) and by McGlynn and van
Kampen (2008 {\it Eur. J. Phys.} {\bf 29} N63-N67). The authors'
misconceptions are recurrent points in the literature. \pagebreak

\noindent {\bf 1. Introduction}

\noindent Recently, Franklin [1] published a thought-provoking and
curious paper on Lorentz contraction and related issues. While his
intention was `to correct students' misconceptions due to
conflicting earlier treatments', I believe that the paper could be
confusing reading for the student. It is a hardly extricable and
certainly challenging mixture of truths, half-truths and erroneous
statements about some basic relativistic concepts and
results.\footnote [1] {I have a vague feeling that perhaps just the
challenge is the concealed essence of [1]. That is, as if Franklin's
true intention was to push to extremes some problematic points in
special relativity and in this way to stir up the student to brood
over them. Considered as a spur, Ref. [1] seems to be an excellent
reading.} Thus Franklin's argument may sound correct to the
unexperienced ear. The situation is aggravated by the circumstance
that some of his contentions are in conformity with interpretations
found in authoritative books on the subject.

In another recent paper, McGlynn and van Kampen [2] contend that the
phenomenon of different values of charge densities in a
current-carrying wire as measured by observers in different inertial
frames, due to relativistic length contractions, is an effect `which
perfectly demonstrates ``the pole in a barn" paradox' at room
velocities. I think, however, that the authors are wrong here,
namely, two distinct aspects of relativistic length contraction are
exemplified in the two situations.

Neither paper is an exception. It is a notorious fact that
understanding relativity is a painful, nay agonizing
process.\footnote [2] {With relativity, we enter the zone of
evolution of the human race in a dramatic way. At a first step
towards the conquest of relativistic mentality, one strikes the hard
wall of everyday language. As Schr\"{o}dinger pointed out `...
everyday language is prejudicial in that it is so thoroughly imbued
with the notion of time - you cannot use a verb ({\it verbum},
``the" word, Germ. {\it Zeitwort}) without using it in one or the
other tense. ... [Special Relativity] ... meant the dethronement of
time as a rigid tyrant imposed on us from outside, a liberation from
the unbreakable rule of ``before and after".' [3] More recently,
Mermin argued that `... to deal with relativity one must either
critically reexamine ordinary language, or abandon it altogether.'
[4] The present author believes that alas it is impossible to
abandon altogether the metalanguage of everyday speech. Physical
meaning is unavoidably blurred by linguistic meaning and {\it vice
versa}.

On the other hand, the word `agony' need not necessarily have a
painful connotation; choosing the title of this paper, I had also in
mind its Greek sense (`a contest for victory'). } We remind the
reader of the old duel between Dingle and Born about the reciprocity
of time dilation [5,6] which aroused a prolonged controversy in the
{\it Nature}.\footnote [3] {In my opinion, Born won. Dingle has made
the same kind of error the student usually makes: two different
quantities are denoted by one and the same symbol and thus confused.
(A compound event that takes place at various spatial points of an
inertial frame K (corresponding to the motion of a clock at constant
velocity $\pmb v$), and has a duration of $1/\sqrt {(1 - v^2/c^2)}$
K-seconds, and a compound event that takes place at one spatial
point of the same K-frame and has the same duration of $1/\sqrt {(1
- v^2/c^2)}$ K-seconds must not be identified; those events are two
distinct straight lines in Minkowski space.)} A natural inference
that the final outcomes of events must be the same with respect to
two inertial observers cannot be generalized to two infinite
continuous sequences of inertial observers, a lesson the present
author has fully learned only very recently [7,8]. Even the meaning
of the Lorentz contraction, which is generally accepted to be the
simplest relativistic phenomenon, is hard to grasp and becomes the
stumbling block in various contexts [9 - 11].\footnote [4] {The most
recent example of this (related to the fact that traps of language
always lurk in special relativity) seems to be the recognition that
what is a well defined wire segment in one inertial frame is no more
a {\it wire segment} as measured in another frame in the framework
of an elementary model of a current-carrying wire [12 - 14].}

This is small wonder. Recall that, as the first physical consequence
of the Lorentz transformations, the student learns that the length
of a rod which is uniformly moving along itself with velocity $\pmb
v$ is reduced by a factor $(1 - v^2/c^2)^{1/2}$ as compared to the
rod's length measured in its rest frame. While the phenomenon has
been dubbed `relativistic length contraction', the student is
immediately warned `but of course nothing at all has happened to the
rod itself'. However, the term `contraction' connotes shrinking
(`cold contracts metals'), shrinking connotes change, and `change'
in physics involves some happening; what happens {\it before} and
{\it after} is measured in one and the same reference frame. Thus,
learning that the rod contracts, yet nothing has happened to it, the
student strikes the hard wall of everyday language.\footnote [5] {A
consequence of a similar relativistic terminological muddle was
described some time ago in Adler's paper `Does mass really depend on
velocity, dad?' [15]. It appears that in the meantime
velocity-dependent mass  disappeared from special relativity
curricula, as it should.} Moreover, in the usual textbook
presentations of special relativity it is explained that `... the
different measures of length are intimately connected with the lack
of absolute simultaneity' [16], or, in the same vein, `... the
contraction, when we observe it, is not a property of matter but
something inherent in the measuring process' [17].

In a recent paper [18], I pointed out that such interpretations
confuse derivation of the phenomenon and its root, thus adding a
conceptual problem to the terminological one. I argued that there is
also a fundamental active aspect of relativistic length contraction:
a rod initially at rest in an inertial frame, after a constant
velocity $\pmb v$ is imparted to it so that it moves freely and
uniformly along itself, is contracted (its length is reduced), all
with respect to that frame; the phenomenon is due to acceleration of
the rod relative to that frame, and is described by the well-known
formula, under the proviso that the acceleration was rest
length-preserving in the final outcome. I inferred that without the
active aspect of length contraction, i. e. without the rest
length-preserving accelerations, there is no special relativity. The
Lorentz transformation, even the formulation of the principle of
relativity, is built on the active aspect. Thus, there is a
dynamical content of the Lorentz transformation.

It appears that the authors of Refs. [1] and [2] were unaware of
[18] or they chose to ignore it.\footnote [6] {Professor Franklin
has kindly informed me that he had not known of my paper [18], but
he does not think `it would be more appropriate to refer to it than
to any of a great number of papers written about the Lorentz
transformation over a span of 120 years.'} Probably my argument was
not perspicuous enough. In the present paper I will point out what I
find to be weak points in Refs. [1] and [2], and elsewhere. That
would perhaps sharpen my argument of [18] and, hopefully, save the
student of relativity some time and effort.\footnote [7] {The
ubiquitous worm of doubt reminds me of the possibility that {\it my}
analysis of the length contraction phenomenon was wrong. However,
brooding over the issue and brooding again, I was unable  to find
any flaws in my argument [18].}

\medskip

\noindent {\bf 2. The length of a moving rod revisited}

\medskip

\noindent In this section, for the convenience of the reader I will
briefly summarize main conclusions of [18], slightly improving
terminology.

Consider two inertial reference frames $S$ and $S'$ in standard
configuration, $S'$ is uniformly moving at speed $v$ along the
common positive $x, x'$-axes, and the $y$- and $z$-axis of $S$ are
parallel to the $y'$- and $z'$-axis of $S'$, respectively. Take a
solid rod parallel to the  $x, x'$-axes, at a permanent rest with
respect to the $S'$ frame, and let $l_0'$ be the length of the rod
as measured in $S'$ by a given measuring rod also at rest in $S'$.
What is the length of the rod as measured in $S$ employing the same
measuring rod which is now at rest in $S$?

Following Einstein's prescription for ascertaining `the length of a
uniformly moving rod ... in the `resting' frame $S$' [19],\footnote
[8] {In [18], I called the prescription Einstein's `very natural
operational definition' of the length of a uniformly moving rod.
However, the term `definition' can be misleading. It could imply
that perhaps some other definition, leading to a different value of
length of a uniformly moving rod, could be legitimately introduced.
(As Dieks pointed out in his article ``The `reality' of the Lorentz
contraction" the term `definition' `... possesses the connotations
of arbitrariness and conventionality.') As far as I understand
special relativity, the uniformly moving rod has no other length
than the part of a (stationary) straight line taken up
instantaneously by the rod, all relative to $S$.} by using the
Lorentz transformation, one deduces in the well-known way that the
length of our uniformly moving rod as measured in $S$, $l_v$, is
given by

\begin {equation}
l_v = l_0'\sqrt{1 - v^2/c^2}\, .
\end {equation}

\noindent The phenomenon expressed by equation (1), that one and the
same rod has different lengths $l_v$ and $l_0'$ as measured in the
$S$ and $S'$ frames, respectively, where $S'$ is the rest frame of
the rod, in what follows I will call the relativistic length
reduction.\footnote [9] {Equation (1) is usually called
`relativistic length contraction' or `conventional length
contraction' [20]. The term `reduction' seems to be more proper
here, as being perhaps more neutral, than `contraction' (which
connotes shrinking, as was pointed out above). I am grateful to
Professor Giuliano Boella for stimulating correspondence concerning
this terminological point.

Of course, the content of equation (1) is not exhausted by the
length of a moving rod, cf footnote 6 of [18].

It should be noted that the term `one and the same rod' above has a
peculiar, special relativistic meaning: it connotes that no action
was exerted upon the rod by a mere different choice of inertial
reference frame (observer), and yet the rod does not have the same
length in the various frames (corresponding to various
cross-sections of the world-strip of the rod, cf Appendix A). Again,
everyday language is {\it the} problem in special relativity.}

Now in the Relativity Paper [19], Einstein stated that if the same
rod to be measured is at rest in $S$, then `according to the
principle of relativity' its length as measured in $S$, $l_0$, must
be equal $l_0'$,

\begin {equation}
l_0 = l_0'\, ,
\end {equation}

\noindent employing of course the same measuring rod as in the
earlier measurements.\footnote [10] {One should make a clear
distinction between the rest length Lorentz-invariance, which is a
truism, and the rest length frame-independence, which is a
fundamental physical assumption. Namely, if a rod is uniformly
moving along itself with respect to an inertial frame, making a
Lorentz boost to its rest frame $S^*$, one can measure its rest
length $l_0^*$. The quantity $l_0^*$ is {\it a fortiori} Lorentz
invariant. This of course means, on the basis of equation (1), that
$l_0^* = l_{v_1}/\sqrt{1 - v_1^2/c^2} = l_{v_2}/\sqrt{1 -
v_2^2/c^2}$, etc, where $v_1, v_2, ...$, are speeds of the rod
relative to inertial frames $S_1, S_2, ...$, respectively (all the
frames being in standard configuration with $S^*$). On the other
hand, Einstein's assumption (2) has quite a different meaning, which
was perhaps most clearly expressed by Resnick ([21], p 93): `The
{\it rest length} of a rod is an absolute quantity, the same for all
inertial observers: If a given rod is measured by different inertial
observers by bringing the rod to rest in their respective frames,
each will measure the same length.' This rest length
frame-independence could be also termed the absolute
Lorentz-invariance of rest length.} Equations (1) and (2) imply

\begin {equation}
l_v = l_0\sqrt{1 - v^2/c^2}\, .
\end {equation}

Equation (3) relates the length $l_0$ of the rod at rest to its
length $l_v$ when it is in uniform motion along itself at the speed
$v$, all with respect to the inertial frame $S$.\footnote [11]
{While equation (3) is an obvious consequence of eqs. (1) and (2),
one point should be stressed. Namely, according to Einstein, a rod
at rest with its axis lying along the $x$-axis, having the length
$l_0$, after `a uniform motion of parallel translation (with
velocity $v$) along the $x$-axis ... is imparted to the rod' will
have the length $l_v$ given by equation (3), all with respect to the
same inertial frame $S$ [19]. This is so regardless of the way the
speed $v$ was imparted to the rod (Einstein made no restrictions).
Thus, according to Einstein, an {\it arbitrary} acceleration of an
arbitrary rod, starting from rest, with the only proviso that the
acceleration leads eventually to a uniform (unconstrained) motion of
the rod along its length, in a persistent internal state, does not
(eventually) change the rest length of the rod.} The phenomenon
described by equation (3) in what follows I will call the
relativistic FitzGerald-Lorentz contraction, or shortly the
FitzGerald-Lorentz contraction, as I did in [18].\footnote [12] {It
should be pointed out that the FitzGerald-Lorentz contraction does
not have the same content as the original idea put forth by
FitzGerald and Lorentz long time ago, as will be explained in
subsection 3.4 below.} Obviously, in this case there is a change of
the rod relative to $S$ (its length has changed); the change is due
to acceleration of the rod from rest to the state of uniform motion.
This is contrary to the situation described by equation (1), in
which case there is no change of the rod in the standard physicists'
sense of the word (involving alterations in the rod with time in one
and the same {\it inertial} reference frame); only the frame $S$
world-map is substituted for the frame $S'$ world-map, without
exerting any action upon the rod itself. It seems natural to call
the content of eqs. (1) and (3) a passive and an active
interpretation of relativistic length contraction,
respectively.\footnote [13] {One important point should be stressed
here in relation to the passive interpretation of length
contraction. Throughout the present article I insisted that no
action was exerted upon the rod by a mere different choice of
inertial reference frame (and that consequently there is no change
of the rod in the standard physicists' sense of the word). This is
of course true. It should be noted, however, that even in the case
of the relativistic length reduction there is a change of the rod in
a certain physically reasonable sense. This important point, whose
neglect makes the content of equation (1) hard to grasp, will be
discussed in some detail in Appendix B. The essence is that
something has happened to the rod even in the passive interpretation
of length contraction.}

Now, there is a well-known tradition, originated by Einstein [19],
to present relativistic length contraction as a purely kinematical
effect. Thus, in the usual textbook presentations of special
relativity, the active interpretation of length contraction is
either neglected or introduced tacitly.\footnote [14] {For example,
after a brief discussion of the relativistic length reduction,
equation (1), as a kind of a {\it velocity perspective} effect (`but
of course nothing at all has happened to the rod itself'), Rindler
[22] stated that the phenomenon `is no ``illusion'': it is real and,
in principle, usable.' However, giving an argument for the last
statement, as is clear from the context, he was tacitly assuming
equation (2), i. e. he had tacitly passed from the passive to the
active interpretation of length contraction. Rindler is no
exception. For some mysterious reason, the pride of place has been
given to the passive interpretation by various authors including
Einstein, Born, Pauli. The heuristic level of special relativity,
`helping us to recognize a great miracle of the world' [18], seems
to be usually kept {\it sub rosa}.} However, without the active
interpretation there is no special relativity as a physical theory
which `asserts definite properties of real bodies'. This is clear
from Einstein's definition of two inertial frames in standard
configuration (which conceptually precedes the formulation of the
principle of (special) relativity and a derivation of the Lorentz
transformation and which, as far as I am aware, cannot be replaced
by another definition), and from a related Einstein's assumption of
`the boostability of rulers and clocks' [23], made explicit by Born
[24], cf footnotes 4 and 12 of [18].\footnote [15] {Current textbook
literature seems to imply that two inertial frames in standard
configuration can be introduced simply by {\it fiat}, which is in my
opinion incorrect.} As was pointed out above, the active
interpretation involves changes.

Thus, if there were no change in the (macroscopic) object in special
relativity, then special relativity would not exist as a valid
physical theory (it could not even be formulated). However, changes
which appear in special relativity may have curious properties,
requiring a thorough reexamination of everyday language. As was
pointed out in [18], the FitzGerald-Lorentz contraction described by
equation (3) possesses the following peculiarity: a free rod in
uniform motion along its length is contracted (shrunk) with respect
to the $S$ frame and yet it is perfectly relaxed (with no stress
relative to both $S$ and $S'$ frames), the contraction being its
natural state when it is in that state of motion (all this under the
proviso that the rod was unstressed when initially it was at rest in
$S$). Also, contrary to what was sometimes stated in the literature,
the contraction is not due to the relative motion of a body; it is
due to acceleration (or deceleration, in the reciprocal case of
elongation) of the body relative to an inertial frame.

One last point. My key inference in [18] seems to be that a weaker
assumption than Einstein's original `boostability of rulers and
clocks' is sufficient for foundation of special relativity. The
weaker assumption, which I called in [18] `the universal
boostability assumption', states that it is possible to boost a
measuring rod  or clock in a way which leaves their measuring
capacity untouched.\footnote [16] {Einstein took it for granted that
the measuring capacity of a measuring rod  or clock would remain
untouched under {\it arbitrary} boosts, cf footnote 12 of [18].} As
far as I am aware, this implies that the rest length of a rod need
not be preserved under arbitrary boosts. There is no guarantee of
the absolute Lorentz-invariance of rest length.

\medskip

\noindent {\bf 3. Some weak points in Refs. [1] and [2]}

\medskip

\noindent {\bf 3.1. There is no change in the object in special
relativity}

\medskip

\noindent Franklin based his discussion of length contraction in
special relativity on the following premise: In special relativity,
there is no change in the object. It is only the reference frame
that is changed from $S$ to $S'$. Now since that premise runs as a
common thread through various authoritative discussions of the
topic, it perhaps deserves some clarification.

As a representative example, I choose the famous book {\it
Einstein's Theory of Relativity} by Max Born [24]. In a section
under catchy title {\it Appearance and Reality}, Born pointed out
that some opponents of special relativity assert that Einstein's
theory implies `... a {\it violation of the causal law}. For if one
and the same measuring rod, as judged from the system $S$, has a
different length according to its being at rest in $S$ or moving
relative to $S$, then, so these people say, there must be a cause
for this change. But Einstein's theory gives no cause; rather it
states that the contraction occurs by itself, that it is an
accompanying circumstance of the fact of motion.' Born defended
special relativity by arguing that the opponents have `... a too
limited view of the concept ``change"'. He explained that `the
standpoint of Einstein's theory about the nature of the contraction
is as follows: A material rod is physically not a spatial thing but
a space-time configuration. Every point of the rod exists at this
moment, at the next, and still at the next, and so on, at every
moment of time. The adequate picture of the rod under consideration
(one-dimensional in space) is thus not a section of the $x$-axis but
rather a strip of the $x,ct$-plane [parallel to the $ct$-axis] ...
The ``contraction" does not affect the strip at all but rather a
section cut out of the [corresponding] $x$-axis. It is, however,
only the strip as a manifold of world points (events) which has
physical reality, and not the cross-section. Thus the contraction is
only a consequence of our way of regarding things and is not a
change of physical reality. Hence it does not come within the scope
of the concepts of cause and effect.'

It is clear that Franklin's premise concurs with Born's explanation:
there is no change in the object in special relativity. It is also
clear that the authors would be right if their arguments referred
only to the relativistic length reduction, described by equation
(1). However, if equation (1) were the whole contents of length
contraction, i. e. if there were no change in the object in special
relativity, then special relativity would not exist as a physical
theory, as is pointed out in the preceding section (cf also [18]).

To do justice to Born, it should be noted that in the first part of
section {\it Appearance and Reality} of [24], introducing his
`principle of the physical identity of the units of measure', he
essentially argued for a change in the object in special relativity
(namely, that equation (1) and assumption (2) imply the physical
validity of equation (3)).\footnote [17] {This point will be
discussed in some detail in Appendix A.} Unfortunately, in the
sequel he confused the (relativistic) FitzGerald-Lorentz contraction
(where change is obvious) with the relativistic length reduction
(where there is no change in the usual physicists' sense of the
word), ascribing properties of the second phenomenon to the first
one. Thus his defense of special relativity failed.\footnote [18]
{In {\it Appearance and Reality}, Born switched several times,
tacitly and obviously unconsciously, between the active and the
passive interpretation of length contraction, using the same term
`contraction' for both phenomena, and confusing their meanings. This
section of Born's book ([24], pp 251-62) is perhaps a perfect
example of how unavoidably terminological confusion leads to
conceptual confusion. This confusion seems to be commonplace in the
literature. Thus Pauli in his discussion of the Lorentz contraction
[25] confused eqs. (1) and (3), obviously assuming tacitly equation
(2). While dealing only with the relativistic length reduction, he
ended the discussion with the following query which clearly refers
to the active aspect of the phenomenon: `Should one then ...
completely abandon any attempt to explain the Lorentz contraction
atomistically? We think that the answer to this question should be
No.' ([25], pp 11-15) Needless to say, it would be much simpler to
explain equation (3) atomistically, than equation (1).}

\medskip

\noindent {\bf 3.2. There is only one length: the `rest frame
length'}

\medskip

\noindent As was pointed out above, Franklin's premise in [1] is
that there is no change in the object in special relativity.
However, since properties like length undergo changes, the author
cut the Gordian knot as follows: the Einstein length of a moving
object is not a physical attribute of the object! Only its `rest
frame length' is a physically reasonable attribute - length of the
object.\footnote [19] {It seems that Franklin thus introduced a
novel and radical interpretation of special relativity which, I
think, can be summarized by the following manifesto: If some
consequences of special relativity are surprising and hard to
understand, they should be proclaimed devoid of physical sense.}
Moreover, discussing the relativistic length reduction, equation
(1), as a {\it velocity perspective} effect, he inferred that `the
``shortening" of a stick that is rotated in four dimensions by a
Lorentz transformation is ... illusory.' However, stating that the
relativistic length reduction is an illusion would represent a
falsification of special relativity.

Now one of Franklin's starting assertions, that `the measured length
of a moving object depends on the ``particular way" in which it is
measured', is perfectly correct. Indeed, one and the same moving
object may have various measured lengths, depending on which {\it
definition} (i. e. which procedure of measurement) of the length of
a moving object is being used, all with respect to one and the same
reference frame. However, for some reason Franklin ignored the
fundamental fact that according to Einstein's special relativity the
moving object has only one length in the (stationary) frame $S$,
that obeying equation (1); it is the only physical reality
(world-map) for the $S$-observer. A photograph of a (small) moving
object would indeed be identical to a photograph of an object that
is somewhat rotated, but of the same shape and dimensions as
compared with the moving object in its rest frame, under the proviso
that the rotated object is at rest, as Franklin recalled. However,
as is well known, that inference is reached assuming special
relativity which means, {\it inter alia}, that the moving object has
only one length with respect to $S$, i. e. that the relativistic
length reduction had taken place (cf [17], pp 150-2, [16], pp 163-8,
[26], [27]).\footnote [20] {Recall that at the end of his paper [28]
Terrell stressed that none of his statements there `should be
construed as casting any doubt on either the observability or the
reality of the Lorentz contraction [i. e. the relativistic length
reduction], as all the results given are derived from the special
theory of relativity.' It is perhaps worthwhile to mention here
that, analyzing in 1905 how the shape of a body depends on the
reference frame in which it is measured, Einstein occasionally used
the verb `betrachten'. This German verb has two meanings: first, to
observe, to see, and second, to consider, depending on the context.
Various English translators of the Relativity Paper seem to be
unanimous that Einstein used `betrachten' in the first sense. (The
present author shares this point of  view.) Wind hindsight, we know
today that Einstein (and translators) should have been using {\it to
consider}, or perhaps better {\it to measure} (`messen', in German),
instead of {\it to observe}, {\it to view}. (Here of course {\it to
measure} is used in the sense of Einstein's `operational
definition'.) The moral of the story seems to have been known to
Democritus: things are not found where their picture is.} What
Franklin characterizes as `the belief that a moving object has a
different length', is the only physical reality for the
$S$-observer; the `belief' is obviously built into the standard
Lorentz transformations as $x'\sqrt{1 - v^2/c^2} = x - vt$, i. e. in
the more familiar form

$$
x' = \gamma_v (x - vt)\, .
$$

\noindent On the other hand, it is clear that Franklin in [1] does
assume the validity of the Lorentz transformations. Thus his
argument appears to be self-contradictory. Contrary to Franklin's
statement, one and the same moving rod has infinitely many lengths
in infinitely many inertial frames in the standard configuration
with its rest frame $S'$, respectively. (Note, however, that the
different lengths are, {\it in a certain sense}, due to {\it
changes} of the rod, cf Appendix B.) According to special
relativity, none of the lengths is less or more {\it physically}
real than the rod's rest length $l_0'$; each inertial observer
possesses her or his perfectly legitimate physical reality.\footnote
[21] {It could be perhaps somewhat misleading to state, as French
does, that Einstein's length of a moving rod in the stationary frame
$S$, obeying equation (1), `does refer to measurements of a
particular kind ...' ([17], p 152). Einstein's `operational
definition' is not so much a {\it measurement} of a particular kind
but rather a perfectly classical explanation of what is the length
of a moving rod (regardless of how it moves), whose only peculiarity
is that one instant of the $S$-time should be understood according
to special relativity. Namely, the Einstein length is the length of
a segment of a stationary line taken up instantaneously by the
moving rod, all with respect to $S$. What else on earth could be the
length of a moving object? Measuring that {\it stationary} length
would hardly be a measurement of a particular kind. However, to
ascertain the stationary line segment would require, e. g., taking a
photograph of the moving rod and of course a clever interpretation
of the photograph [27]. }

Another Franklin's basic assertion is that two different inertial
frames are required `in order to compare the measured length of a
moving object to its measured length in a system in which it is not
moving'. This is of course true in the case of the relativistic
length reduction, described by equation (1). However, if that were
the whole content of special relativity, then it would not exist as
a physical theory. As was pointed out in [18] and also above, the
foundation of special relativity requires the rest length-preserving
accelerations (and also, more generally, it requires the rest
properties-preserving accelerations (cf [29, 30])). In the case of
such {\it gentle} accelerations (which are {\it sine qua non} for
special relativity) one inertial frame would be enough for a
comparison of the two lengths.

\medskip

\noindent {\bf 3.3. Who contends stresses can be induced by Lorentz
contraction}

\medskip

\noindent Franklin stated in [1] that Lorentz contraction (by which
he obviously meant the relativistic length reduction, equation (1))
could not induce strains and stresses. He illustrated this with a
simple example of a brittle wine glass at rest on a table, pointing
out that moving past the wine glass at constant velocity (and
looking at it) could not shatter the wine glass. This is of course
true: an object at permanent rest and perfectly relaxed in the $S'$
frame, is perfectly relaxed also relative to the $S$ frame (no
action was exerted upon the object in the change of the inertial
reference frame from $S'$ to $S$).\footnote [22] {Moreover, as was
pointed out in [18], with the rest length-preserving accelerations
the object is perfectly relaxed relative to $S$, while being
contracted (shrunk) relative to $S$.}

Now Franklin also stated that Refs. [31-34] contended that stresses
and strains could be induced by Lorentz contraction. However, as far
as I can see, there is no hint of such a contention in FitzGerald's
five-sentence letter to {\it Science}, where he had suggested a
hypothesis that `the length of material bodies changes, according as
they are moving through the ether or across it, by an amount
depending on the square of the ratio of their velocity to that of
light' ([31], cf also [35]).\footnote [23] {In my opinion, no
distortion or deformation, and thus no stresses, are implied in the
following FitzGerald's sentence in [31]: `We know that electric
forces are affected by the motion of the electrified bodies relative
to the ether, and it seems a not improbable supposition that the
molecular forces are affected by the motion, and that the size of a
body alters consequently.' (It seems to me that the only way of
finding stresses in the FitzGerald sentence would be to read
Lorentz's ideas into it.) However, I agree with Brown ([23], p 51)
that the FitzGerald supposition was prompted by Heaviside's result
for the electromagnetic field of a point charge in uniform motion
relative to the ether.} Also, despite appearances, no such
contention is the {\it essence} of Refs. [33], [34]. Namely, Bell
clearly stated that `... the artificial prevention of the natural
contraction imposes intolerable stress' [34], where by `the natural
contraction' he obviously meant the {\it relativistic}
FitzGerald-Lorentz contraction, described by equation (3). (Nowhere
in [34] Bell stated that `the natural contraction' itself induces
stresses.) On the other hand, it is true that Dewan and Beran [33]
described their {\it Gedankenexperiment} as a demonstration `that
relativistic contraction can introduce stress effects in a moving
body'. However, the authors were somewhat sloppy in their
wording\footnote [24] {For example, their formulation that {\it the
Lorentz transformation} `... implies that a fast moving object
contracts in the direction of its velocity' is in the general case
incorrect. Namely, the relativistic length {\it reduction} involves
no {\it change} in the object, as was pointed out above. Similarly,
Bell [34] spoke about `systematic {\it distortion} of the field of
fast particles' (italics added by D. V. R.) as compared with the
spherically symmetrical Coulomb field of a charge at rest. However,
I am convinced it is simply a bad wording.}, as is often the case in
discussions dealing with special relativity; I think it is clear
from the contents of [33] that their {\it intended meaning} is
perfectly summarized by the above quotation from Bell. Thus it seems
that only Lorentz ([36], pp 5-7, 21-23, 27-28) spoke explicitly
about deformation (and thus about stresses) of a body in connection
with his `by no means far-fetched' hypothesis that if to a system
$\Sigma'$ of particles in the equilibrium configuration, at rest
relative to the ether, `the velocity $\pmb v = v\pmb {\hat{x}}$ is
imparted, it will {\it of itself} change into the system $\Sigma$
[which is got from $\Sigma'$ by the deformation $(\frac {1}{\beta
l}, \frac {1}{l}, \frac {1}{l})$, where $\beta = (1 -
v^2/c^2)^{-1/2}$ and $l$ is a numerical factor allowing for a change
in the $y$ and $z$ directions]. In other terms, the translation will
{\it produce} the deformation $(\frac {1}{\beta l}, \frac {1}{l},
\frac {1}{l})$.'\footnote [25] {Perhaps the Lorentz formulation that
bodies `have their dimensions changed by the effect of translation'
prompted Minkowski to characterize the hypothesis as sounding
`extremely fantastical, for the contraction ... [is to be looked
upon] simply as a gift from above, - as an accompanying circumstance
of the circumstance of the motion.' ([36], p 81). With the benefit
of hindsight, and taking the liberty of rectifying FitzGerald and
Lorentz, I believe that both eminent physicists were victims of the
traps of ordinary language: concerning their statement that bodies
are changed by their translational motion relative to the ether, I
think that their intended meaning was that bodies are changed by
their acceleration relative to the ether from rest until reaching a
steady velocity.} (Note that Lorentz subsequently showed that $l =
1$ ([36], p 27).)

\medskip

\noindent {\bf 3.4. The original FitzGerald-Lorentz contraction and
its relativistic counterpart}

\medskip

\noindent Franklin's starting statement in [1] is that `Lorentz
contraction [introduced by FitzGerald and Lorentz] is not what
actually occurs for a moving body in special relativity'. While that
statement is certainly correct, it seems that the author in the
sequel ignored the fact that there is a perfectly legitimate
FitzGerald-Lorentz contraction in special relativity. This point
perhaps needs some clarification.

As is well known, FitzGerald and Lorentz introduced the contraction
(shrinking) of bodies in motion relative to the ether. Thus a rod at
rest on the earth may be contracted, depending on the direction of
its motion through the ether, as compared (measured) with the same
rod at rest in the ether. However, in the world-map of the
FitzGerald and Lorentz, a rod at rest in the ether is not contracted
in comparison with an identical rod which is brought to rest
relative to the earth; rather, the former rod may be elongated as
compared with the latter. In this sense, the contraction introduced
by FitzGerald and Lorentz is absolute, there is no reciprocity in
it.\footnote [26] {This is contrary to the relativistic length
reduction, which is reciprocal. Note, however, that the relativistic
FitzGerald-Lorentz contraction, which refers to one and the same
reference frame, is not reciprocal too.} As Franklin pointed out,
applying this originally introduced FitzGerald-Lorentz contraction
to a variant of the Michelson-Morley apparatus would lead to a {\it
positive} result (cf [37], p 236).\footnote [27] {Note, however,
that that conclusion is based on the premise that the velocity of
light is the same in all directions only in the ether frame,
contrary to Franklin's assertion.}

Now a clear distinction should be made between the original
FitzGerald-Lorentz contraction and its relativistic counterpart
(which is also called - and justly so - the FitzGerald-Lorentz
contraction). Namely, what the two conceptions have in common is
shrinking (which I think is due to acceleration); however, contrary
to the former, the latter actually occurs for an object to which a
constant velocity is imparted in {\it any} inertial frame (under the
proviso of the rest length-preserving accelerations). (Recall that
there is no ether frame in special relativity simply by virtue of
Ockham's razor (for an interesting argument cf [38]).\footnote [28]
{By the way, we remind the reader that the ether played a rather
subtle part for the {\it fin de si\` ecle} electrodynamicists (cf
[39], and [40], pp 176-7).})

\medskip

\noindent {\bf 3.5. The Bell spaceship paradox}

\medskip

\noindent Section 2 entitled `The Bell spaceship paradox' seems to
be the most mischievous part of [1]. While Franklin asserts that he
presents `the nexus of the Bell spaceship paradox as originally
presented by John Bell' [34], actually this is not so. Namely, Bell
took into account the Evett and Wangsness correction of the original
Dewan-Beran formulation of the problem [41]. Thus, instead of
connecting the tail of the front spaceship (R) and the nose of the
back spaceship (L) as is supposed in [1], [33], in the Bell
formulation a thread connects the {\it corresponding points} of
ships.\footnote [29] {This point is specially clear in the `mild'
variant of the problem, in which at an instant of the $S$ time the
ships' acceleration ceases and they coast with the same constant
velocity, as measured in the $S$ frame [11]. Namely, assuming the
rest length-preserving accelerations of {\it ships} in the final
outcome, ships will eventually FitzGerald-Lorentz contract according
to equation (3), and thus the final distance between the tail of R
and the nose of L will be greater than their initial distance, all
with respect to $S$. It is perhaps worthwhile here to clarify the
standard assumption that the thread connecting ships in no way
affects the {\it motion} of ships. Namely, this does {\it not} mean
that the thread does not affect {\it ships} (it does, cf [42]);
instead that means that the work programmes of the ships' motors are
being constantly re-adjusted so as to provide ships having identical
accelerations with respect to $S$. } Moreover, Franklin's analysis
of his version of Dewan-Beran-Bell's problem (which obviously refers
to the `tough' variant of the problem\footnote [30] {This means that
ships have identical accelerations $a(t)$ in the positive
$x$-direction starting simultaneously from rest, all with respect to
$S$; in the general case, the ships' acceleration never ceases.}) is
basically incorrect: contrary to Franklin's repeated statements,
there is no common rest frame $S'$ for both ships (`even for
continually accelerated spaceships'), as is clear from the
corresponding Minkowski diagram. (Events that are simultaneous in
the $S$ frame are not simultaneous with respect to any other frame,
and {\it vice versa}.) Consequently, there is no rest frame distance
between ships (there is no frame in which both ships are {\it
simultaneously} at rest, except of course the $S$ frame at $t =
0$)\footnote [31] {Thus Franklin's contention in section 1 of [1]
that `it is only the rest frame length of an object that relates to
strains and stresses on the object' is in the general case wrong.}
and equations (2)-(6) in [1] are meaningless, for continually
accelerated ships. (They are not incorrect, they are meaningless,
since there is no the $S'$ frame.\footnote [32] {The correct
distances between the {\it corresponding points} of ships R and L,
when the points are performing identical hyperbolic motions relative
to $S$, as measured in instantaneous rest {\it frames} of R and L at
the same instant of their proper time $\tau$ are given, e. g., in
[8]. Note that one and the same inertial frame first becomes the
instantaneous rest frame of R and subsequently (with respect to {\it
that} frame) it becomes the instantaneous rest frame of L [8].})

Eventually, Franklin's resolution of the Bell spaceship paradox `as
no paradox' is hard to fathom. It is of course true that special
relativity allows no difference in any measurement of two equal
lengths such as the distance between ships and the length of the
thread between them. However, the two distances are not of the same
sort in the following sense. Consider, for simplicity, the `mild'
variant of the problem when all transient effects have died out and
a steady velocity of ships is reached in $S$. Then to the former
distance the relativistic length reduction applies, whereas to the
latter (the length of the thread) both the length reduction and a
stretching above the natural relativistic FitzGerald-Lorentz
contraction of the thread apply (under the proviso that the thread
remained {\it unbroken} and under the proviso of course that special
relativity is valid).\footnote [33] {It should be pointed out that
the conclusion `a stretching above the natural FitzGerald-Lorentz
contraction of the thread applies' in the mild variant of the
problem is based on the tacit assumption that releasing the thread
ends from ships in the final rest frame $S'$ would lead to the
thread's shrinking in $S'$ to its initial rest length (the length it
had in $S$ before accelerations started), or in other words that the
thread is perfectly elastic. Without that simplifying assumption the
analysis of Dewan-Beran-Bell's problem becomes tricky in $S$. Thus,
without that simplifying assumption Bell's resolution of the paradox
in $S$ is oversimplified. However, at a sufficiently high speed the
thread would certainly break regardless of its elasticity since,
according to special relativity, its length in $S'$  would tend to
infinity when $v \rightarrow c$. The same conclusion is reached in
the $S$ frame, taking into account that the thread's natural
(FitzGerald-Lorentz contracted) length when it is in uniform motion
would tend to zero when $v \rightarrow c$. (The appearance of [11]
stimulated heated discussions on some internet physics forums and
several published [7,8,43] and unpublished papers on the topic. An
anonymous Russian author, a philosopher by profession, remarked that
`it would not be a big harm if a philosopher added into the barrel
of professional physicists' honey a teaspoon of philosophical tar'.
The present footnote is prompted by the author's comment that
special relativity {\it alone} does not imply the thread would break
due to `the artificial prevention of the natural contraction.')}
While our Galilean instincts would expect the thread never breaks in
$S$ (why should it?), it must break at a sufficiently high speed if
special relativity is valid, and this is the core of the paradox.

\medskip

\noindent {\bf 3.6. Rigid body motion in special relativity}

\medskip

\noindent Franklin starts the last section of [1] entitled `Rigid
body motion in special relativity' by pointing out that in the
motion described by Bell and by himself, the acceleration of each
spaceship is the same at equal times in $S$. He then contends `this
also corresponds to each [spaceship] having the same acceleration
$a'$ in their [mutual] instantaneous rest system ... if their rest
system acceleration is constant in time.' However, as was noted
above, there is no common instantaneous rest frame for both ships;
instead, ships' accelerations are constant in their respective
instantaneous rest frames. In the same way, Franklin's next argument
that `from the preceding paragraph we see that keeping lengths
constant in the rest system requires different rest frame
accelerations for different parts of a rigid body' is inconclusive:
there is no {\it mutual} instantaneous rest frame in
Dewan-Beran-Bell's problem.

It should be stressed that Franklin's subsequent analysis of the
motion of ships in the case of their constant but different rest
frame accelerations, so as to keep the distance between ships
constant in their mutual rest system, is exact and instructive.
There is only one terminological point where I disagree with
Franklin. Namely, his formulation `rigid body motion', which
reflects the concept originally introduced by Born [44], should be
replaced by a more proper formulation `rigidly moving body' (cf
[25], pp 130-2, [40]).

\medskip

\noindent {\bf 3.7. Linking electrical current and the pole in a
barn paradox}

\medskip

\noindent In a recent paper, McGlynn and van Kampen contend that the
phenomenon of alterations in charge densities in a current-carrying
wire as measured by different inertial observers `perfectly
demonstrates ``the pole in a barn" paradox' [2]. However, this is
wrong; the two phenomena exemplify two distinct aspects of
relativistic length contraction, namely, the length reduction and
the FitzGerald-Lorentz contraction, respectively. Since the
confusion appears to be a recurrent point in various contexts [9,
10], and taking into account that it is closely connected with our
preceding considerations, it is perhaps worthwhile to briefly
discuss McGlynn and van Kampen's contention.

The standard textbook derivations of the magnetic force that acts on
a moving charge $q$ via special relativity consider the case of an
infinite straight wire at rest in the `laboratory' frame ([17], [45,
46]). The wire is modelled as consisting of two superposed lines of
charge: one moving (that of free electrons moving at drift speed
$v_d$) and the other, which has an equal but opposite charge
density, at rest (that of fixed positive ions). Thus, the wire is
taken to be electrically neutral in the laboratory frame ($S$),
which implies that the distance between adjacent ions equals the
(mean) distance between adjacent electrons in $S$. Then by applying
the proper relativistic length reduction formulae ({\it mutatis
mutandis } in equation (1)) to those distances, the corresponding
charge densities in the rest frame of the moving charge $q$ are
found. Eventually, following the well  known relativistic path,
making (tacitly) use of the happy circumstance that the Lorentz
force is a {\it pure} relativistic force\footnote [34] {This
implies, {\it inter alia}, that the Lorentz force transforms in the
same way as $\frac {d}{dt}(m\pmb u \gamma_u)$, where $m$ is {\it a
time-independent Lorentz scalar}, and $\pmb u$ is the instantaneous
velocity of a particle.} ([22], [16], [47], cf also [48], p 129,
[49]), the desired result for the magnetic force is obtained. Note
that in the above scene-setting no contractions are involved in the
$S$ frame.\footnote [35] {As Zapolsky [9] pointed out, Feynman {\it
et al} [45] and Purcell [46] introduce electrical neutrality of the
current-carrying wire in the laboratory frame $S$ by {\it fiat}. On
the other hand, Zapolsky offered an explanation for the neutrality
arguing that a steady current is established in a conducting wire
(which is electrically neutral in $S$ when no current flows in it)
turning on a constant electric field oriented along the wire {\it
simultaneously} (relative to $S$) at all points of the {\it
infinite} wire, etc. That explanation seems to be implicit in a
related quotation from French ([17], p 259):

`It is important to note that {\it no} contractions are involved
from the standpoint of the laboratory frame, but only from the
standpoint of a frame moving relative to the laboratory. The only
difference between a wire carrying a current and a wire not carrying
a current is the existence of a drift velocity for the electrons.
The mean distance between the electrons remains unaffected as
measured in the laboratory frame.'

In the present paper, I refrain from entering into tricky problems
of how is a current established in a conducting wire, and in what
frame is a current-carrying conductor neutral. For purposes of
teaching the relationship between electricity and magnetism via
special relativity, it seems reasonable to introduce the neutrality
of the wire in $S$ by {\it fiat}. Thus {\it a fortiori} no
contractions are involved in the $S$ frame.} The same scene was used
in [2], except for the fact that McGlynn and van Kampen confined
their attention to a {\it segment} of the wire in $S$, which is
irrelevant for the present discussion. (Note also that no
contractions are involved (the thread apart) concerning the {\it
distance} between the corresponding points of ships in
Dewan-Beran-Bell's problem {\it in the $S$ frame}.)

The situation is different in the pole in a barn problem ([50], cf
also [22]). Namely, a pole vaulter (who, according to Dewan's
original formulation, lives in `Tompkins' Wonderland' where the
speed of light is low) must speed up his pole from rest in a rest
length-preserving way, so that the FitzGerald-Lorentz contraction
formula (3) applies. In this case obviously there is contraction in
the $S$ frame (which is now the barn frame). Thus, there is a basic
distinction between the two phenomena described above, contrary to
McGlynn and van Kampen's claim in [2].

In more detail, the FitzGerald-Lorentz contraction (and thus also
the corresponding preparatory stage in which the pole acquires its
motion relative to the barn) is essential in the pole in a barn
problem; the contraction makes it possible that the pole in motion
enters (momentarily) the barn (while this was impossible when the
pole was at rest with respect to the barn). Consequently, a change
of the pole with respect to the (inertial) barn frame is essential,
and thus the active aspect of length contraction, expressed by my
eq. (3), is essential in the pole in a barn problem.

On the other hand, we have another story in the phenomenon of
alterations in charge densities in a current carrying wire as
measured by {\it different inertial observers}. First, there is no
contraction wrt the wire (laboratory) frame, in the case of the
steady state assumed by McGlynn and van Kampen [2], Zapolsky [9],
French [17], Feynman {\it et al} [45], Purcell [46]. Second, what is
essential for the phenomenon is {\it the assumed steady state} (the
wire carrying the steady current is electrically neutral in the wire
frame); the preparatory stage (starting from electrically neutral
wire with no current) is irrelevant for the phenomenon. Third, the
alterations in charge densities are found using the relativistic
length reduction formula, my eq. (1). Therefore, the passive aspect
of length contraction is exemplified in the phenomenon.

\medskip

\noindent {\bf Acknowledgments}

\medskip

\noindent I have benefitted a great deal from stimulating and
cordial correspondence with Brian Coleman, Giuliano Boella, Vladimir
Hnizdo, Jerrold Franklin, Dmitry Peregoudov, Paul van Kampen and
Jaykov Foukzon.

\medskip

\noindent {\bf Appendix A}

\medskip

\noindent Assume a rod of unit length at rest in $S$, lying along
the $x$-axis, taking up the segment between the origin and the point
$x = 1$m. The adequate picture of the rod (one-dimensional in space)
is a strip of the $x, ct$-plane, bounded by the $ct$-axis ($x = 0$)
and the line $x = 1$m parallel to it. It is the strip as a manifold
of world points which has {\it objective reality}. At various
instants of the $S$-time, the rod is represented by cross-sections
of the strip parallel to the $x$-axis. In the $S'$ frame, however,
the same rod is represented by cross-sections of the strip parallel
to the corresponding $x'$-axis, at various instants of the
$S'$-time; the length of the rod is $\sqrt{1 - v^2/c^2}$m$'$, as
measured in $S'$. (Recall that the lesser $S'$-length is a {\it
longer} line segment than the $S$-length on the corresponding
Minkowski diagram, due to well-known properties of space calibration
hyperbola $x^2 - c^2t^2 = 1$.) Thus to one and the same objective
reality (the strip) correspond various {\it physical realities}
(cross-sections of the strip parallel to the corresponding spatial
axes), being the world-maps of the {\it same} rod in various
reference frames. In this sense, each inertial frame has its own
physical reality.

Assume now that the velocity $\pmb v = v\pmb {\hat{x}}$ is imparted
to the rod so that it moves uniformly along its length (the
$x$-axis) with respect to $S$, and assume also that the acceleration
was a rest length-preserving one. (This assumption is contained in
Born's `principle of the physical identity of the units of
measure'.) In this case, the corresponding objective reality of the
rod is depicted by a strip of the $x, ct$-plane inclining to the
$ct$-axis, bounded by the $ct'$-axis ($x' = 0$), and the line $x' =
1$m$'$ parallel to it. Cross-sections of the inclined strip parallel
to the $x$-axis are physical reality for the $S$-observer, their
length being of course $\sqrt{1 - v^2/c^2}$m, whereas cross-sections
parallel to the $x'$-axis are physical reality for the
$S'$-observer, their length being 1m$'$ (cf footnote 8 of [18]).

It is clear that there is a change in the object due to acceleration
with respect to the $S$-frame (or, equivalently, due to deceleration
with respect to the $S'$-frame): objective reality (the strip) has
changed; this is so, of course, for all inertial observers.
(However, despite the physical change has happened, the object is
still one and the same object in the sense that it is still a bound
configuration consisting of the same material points.)

In the above argument, Born's term `physical reality' is replaced by
`objective reality'; on the other hand, I used `physical reality' of
an inertial observer as a synonym for Rindler's world-map (cf [22],
and also [18]). Note that, despite appearances, my term `objective
reality' does not necessarily imply a reality which would be
independent of the realm of our perceptions. Note also that my
argument is in accord with that presented by Minkowski in his famous
address `Space and Time' more than a hundred years ago ([36], pp
74-91).

\medskip

\noindent {\bf Appendix B}

\medskip

\noindent Even in the case of the relativistic length reduction,
{\it there is a change of the rod} in the following, physically
reasonable sense. Namely, a {\it change} of inertial frame from $S'$
to $S$, entails essentially the following procedure: it entails
accelerating a reference frame which is an exact copy of $S'$, that
was initially at rest with respect to $S'$, until reaching a steady
velocity with respect to the inertial $S'$, and this in a rest
properties-preserving way. During the acceleration, there is an {\it
inertial force} acting on the rod with respect to the accelerated
frame; acceleration of the rod with respect to the accelerated
(non-inertial) frame is the cause of the rod's shortening with
respect to the (eventually again inertial) frame $S$. This explains
{\it physically}  different lengths that appear in the relativistic
length reduction formula (1).

As far as I know, this important point was not emphasized in the
literature.

\newpage

\noindent {\bf References}

\medskip

\noindent [1] Franklin J 2010 Lorentz contraction, Bell's spaceships
and rigid body motion in special relativity {\it Eur. J. Phys.} {\bf
31} 291-8

\noindent [2] McGlynn E and van Kampen P 2008 A note on linking
electrical current, magnetic fields, charges and the pole in a barn
paradox in special relativity {\it Eur. J. Phys.} {\bf 29} N63-N67

\noindent [3] Schr\"{o}dinger E 1977 {\it What is Life \& Mind and
Matter} (Cambridge: Cambridge UP) pp 158-61

\noindent [4] Mermin N D 1999 Writing Physics

online at http://www.lassp.cornell.edu/\verb
~cew2/KnightLecture.html

\noindent [5] Dingle H 1962 Special theory of relativity {\it
Nature} {\bf 195} 985-6

\noindent [6] Born M 1963 Special theory of relativity {\it Nature}
{\bf 197} 1287

\noindent [7] Peregoudov D V 2009 Comment on `Note on
Dewan-Beran-Bell's spaceship problem' {\it Eur. J. Phys.} {\bf 30}
L3-L5

\noindent [8] Red\v zi\' c D V 2009 Reply to `Comment on ``Note on
Dewan-Beran-Bell's  spaceship problem"' {\it Eur. J. Phys.} {\bf 30}
L7-L9

\noindent [9] Zapolsky H S 1988 On electric fields produced by
steady currents {\it Am. J. Phys.} {\bf 56} 1137-41

\noindent [10] Cavalleri G and Tonni E 2000 Comment on ``\v Cerenkov
effect and the Lorentz contraction" {\it Phys. Rev. A} {\bf 61}
026101-1-2

\noindent [11] Red\v zi\' c D V 2008 Note on Dewan-Beran-Bell's
spaceship problem {\it Eur. J. Phys.} {\bf 29} N11-N19

\noindent [12] van Kampen P 2008 Lorentz contraction and
current-carrying wires {\it Eur. J. Phys.} {\bf 29} 879-83

\noindent [13] Red\v zi\' c D V 2010 Comment on `Lorentz contraction
and current-carrying wires' {\it Eur. J. Phys.} {\bf 31} L25-L27

\noindent [14] van Kampen P 2010 Reply to `Comment on ``Lorentz
contraction and current-carrying wires"' {\it Eur. J. Phys.} {\bf
31} L29-L30

\noindent [15] Adler C G 1987  Does mass really depend on velocity,
dad? {\it Am. J. Phys.} {\bf 55} 739-43

\noindent [16] Rosser W G V 1964 {\it An Introduction to the Theory
of Relativity} (London: Butterworths)

\noindent [17] French A P 1968 {\it Special Relativity} (London:
Nelson)

\noindent [18] Red\v zi\' c D V 2008 Towards disentangling the
meaning of relativistic length contraction {\it Eur. J. Phys.} {\bf
29} 191-201

\noindent [19] Einstein A 1905 Zur Elektrodynamik bewegter K\"
{o}rper {\it Ann. Phys., Lpz.} {\bf 17} 891-921

\noindent [20] Styer D F 2007 How do two moving clocks fall out of
sync? A tale of trucks, threads, and twins {\it Am. J. Phys.} {\bf
75} 805-14

\noindent [21] Resnick R 1968 {\it Introduction to Special
Relativity} (New York: Wiley)

\noindent [22] Rindler W 1991 {\it Introduction to Special
Relativity} 2nd edn (Oxford: Clarendon)

\noindent [23] Brown H R 2005 {\it Physical Relativity: Space-time
Structure from a Dynamical Perspective} (Oxford: Clarendon)

\noindent [24] Born M 1965 {\it Einstein's Theory of Relativity}
(New York: Dover) (revised edition prepared in collaboration with
G\"{u}nther Leibfried and Walter Biem)

\noindent [25] Pauli W 1958 {\it Theory of Relativity} (London:
Pergamon) (reprinted 1981 transl. G Field (New York: Dover))

\noindent [26] Gamow G 1961 Remarks on Lorentz contraction {\it
Proc. Natl. Acad. Sci.} {\bf 47} 728-9

\noindent [27] Kraus U 2008 First-person visualizations of the
special and general theory of relativity {\it Eur. J. Phys.} {\bf
29} 1-13

\noindent [28] Terrell J 1959 Invisibility of the Lorentz
contraction {\it Phys. Rev.} {\bf 116} 1041-5

\noindent [29] Red\v zi\' c D V 2005 Momentum conservation and
Einstein's 1905 {\it Gedankenexperiment} {\it Eur. J. Phys.} {\bf
26} 991-7

\noindent [30] Red\v zi\' c D V 2006 Does $\Delta m = \Delta
E_{\mbox{rest}}/c^2$? {\it Eur. J. Phys.} {\bf 27} 147-57

\noindent [31] FitzGerald G F 1889 The ether and the earth's
atmosphere {\it Science} {\bf 13} 390

\noindent [32] Lorentz H A 1892 De relatieve beweging van de aarde
en den aether {\it Versl. Kon. Akad. Wetensch} {\bf 1} 74-9
(reprinted in translation: Lorentz H A 1937 The relative motion of
the earth and the ether, in {\it Collected Papers} Vol 4 (The Hague:
Nijhoff) pp. 219-223.)

\noindent [33] Dewan E and Beran M 1959 Note on stress effects due
to relativistic contraction {\it Am. J. Phys.} {\bf 27} 517-8

\noindent [34] Bell J S 1976 How to teach special relativity {\it
Prog. Sci. Cult.} {\bf 1} (2) 1-13 (reprinted in Bell J S 1987 {\it
Speakable and Unspeakable in Quantum Mechanics} (Cambridge:
Cambridge UP) pp 67-80)

\noindent [35] Brush S G 1967 Note on the history of the
FitzGerald-Lorentz contraction {\it Isis} {\bf 54} 230-2

\noindent [36] Lorentz H A, Einstein A, Minkowski H and Weyl H 1952
{\it The Principle of Relativity} (New York: Dover)

\noindent [37] Panofsky W K H and Phillips M 1955 {\it Classical
Electricity and Magnetism} 1st edn (Cambridge, MA: Addison-Wesley)

\noindent [38] Mirabelli A 1985 The ether just fades away {\it Am.
J. Phys.} {\bf 53} 493-4

\noindent [39] Torretti R 2006 {\it Can science advance effectively
through philosophical criticism and reflection?}

online at http://philsci-archive.pitt.edu/archive/00002875/

\noindent [40] Miller A I 1981 {\it Albert Einstein's Special Theory
of Relativity: Emergence (1905) and Early Interpretation
(1905-1911)} (Reading, MA: Addison - Wesley)

\noindent [41] Evett A A and Wangsness R K 1960 Note on the
separation of relativistically moving rockets {\it Am. J. Phys.}
{\bf 28} 566

\noindent [42] Cornwell D T 2005 Forces due to contraction on a cord
spanning between two spaceships {\it Europhys. Lett.} {\bf 71}
699-704

\noindent [43] Podosenov S A, Foukzon J and Potapov A A 2009 J.
Bell's problem and research of the electronic clots in linear
collider {\it Nonlinear World} {\bf 7} 612-21 (in Russian)

\noindent [44] Born M 1909 Die Theorie  des starren Elektrons in der
Kinematik des Relativit\"{a}tsprinzips {\it Ann. Phys., Lpz.} {\bf
30} 1-56

\noindent [45] Feynman R P, Leighton R B and Sands M 1975 {\it The
Feynman Lectures on Physics} Vol 2 (Reading, MA: Addison-Wesley) Sec
13-6

\noindent [46] Purcell E M 1985 {\it Electricity and Magnetism} 2nd
edn (New York: McGraw-Hill) pp 190, 192-9

\noindent [47] Jefimenko O D 1996 Derivation of relativistic force
transformation equations from Lorentz force law {\it Am. J. Phys.}
{\bf 64} 618-20

\noindent [48] Red\v zi\' c D V 2002 Electromagnetism of rotating
conductors revisited  {\it Eur. J. Phys.} {\bf 23} 127-134

\noindent [49] Red\v zi\' c D V 2006 {\it Recurrent Topics in
Special Relativity: Seven Essays on the Electrodynamics of Moving
Bodies} (Belgrade: authorial edition) pp 1-2, 11-12, 36

\noindent [50] Dewan E M 1963 Stress effects due to Lorentz
contraction {\it Am. J. Phys.} {\bf 31} 383-6

\end {document}